\documentclass[10pt,twocolumn]{article}

\usepackage[utf8]{inputenc}
\usepackage[T1]{fontenc}
\usepackage{times}
\usepackage[letterpaper,margin=0.75in,columnsep=0.28in]{geometry}
\usepackage{graphicx}
\usepackage{booktabs}
\usepackage{tabularx}
\usepackage{multirow}
\usepackage{amsmath}
\usepackage{amssymb}
\usepackage{url}
\usepackage[hidelinks]{hyperref}
\usepackage{xcolor}
\usepackage{tikz}
\usetikzlibrary{shapes.geometric,arrows.meta,positioning,fit,backgrounds}
\usepackage{caption}
\usepackage{fancyhdr}
\usepackage{enumitem}
\usepackage{abstract}
\usepackage{titlesec}
\usepackage{microtype}
\usepackage{ragged2e}
\usepackage{array}
\definecolor{cvnavy}{HTML}{143A52}
\definecolor{cvgreen}{HTML}{16845F}
\definecolor{cvgray}{HTML}{5C6878}

\titleformat{\section}{\normalfont\large\bfseries\color{cvnavy}}{\thesection}{0.6em}{}
\titleformat{\subsection}{\normalfont\normalsize\bfseries}{\thesubsection}{0.6em}{}
\titlespacing*{\section}{0pt}{10pt}{4pt}
\titlespacing*{\subsection}{0pt}{8pt}{3pt}
\setlist[itemize]{leftmargin=1.1em,itemsep=1pt,topsep=2pt}
\setlist[enumerate]{leftmargin=1.4em,itemsep=1pt,topsep=2pt}

\newcommand{\keywords}[1]{\vspace{4pt}\noindent\textbf{Keywords:} #1}

\title{\vspace{-16pt}\bfseries\color{cvnavy}API Secrets Should Never Become Tokens in the LLM's Vocabulary:\\[3pt]
\large A Threat Analysis of API Credential Handling in LLM Agent Systems and an Empirical Evaluation of a Vault-Mediated Execution Boundary}

\author{
\normalsize
\begin{tabular}{@{}c@{\hspace{3.2em}}c@{\hspace{3.2em}}c@{}}
Patrick Kenney & Hadi Ahmadi & Denis Lusson \\
\small\texttt{pkenney@corvic.ai} & \small\texttt{hadi@corvic.ai} & \small\texttt{denis@corvic.ai} \\[6pt]
\multicolumn{3}{c}{
  \begin{tabular}{@{}c@{\hspace{3.2em}}c@{}}
  Donald Nguyen & Gurbinder Gill \\
  \small\texttt{ddn@corvic.ai} & \small\texttt{gill@corvic.ai} \\
  \end{tabular}
} \\[8pt]
\multicolumn{3}{c}{\emph{Corvic AI Research}} \\
\end{tabular}
}
\date{}

\begin{document}
\twocolumn[
  \begin{@twocolumnfalse}
  \maketitle
  \vspace{-22pt}
  \begin{abstract}\noindent
  \small
  Tool-using large language model (LLM) agents convert credential hygiene from a storage problem into an execution-security problem. When a user pastes an API key into a prompt, or a developer embeds one in a system prompt or tool configuration, the secret crosses from an authentication boundary into a data pipeline where it may be retained in conversation history, logs, memory stores, generated code, and error payloads. Prompt injection and excessive agency can then convert passive disclosure into unauthorized action. This paper (i) formalizes the credential-exposure threat chain for agentic systems, (ii) synthesizes evidence from a platform secret-store incident, vendor-reported secret-sprawl measurement, and OWASP/NIST guidance, (iii) describes a vault-mediated execution architecture in which the model selects a connector identifier while a trusted request boundary supplies authentication, and (iv) reports two controlled black-box experiments against a production implementation of that architecture, Corvic Security Vault. Across 16 evaluated probes spanning seven control domains, every probe met its expected security outcome: an authenticated GitHub API request succeeded while the credential remained absent from process environment values, caller-visible request headers, tested filesystem locations, three third-party echo services, and two unrelated API origins; both cloud instance-metadata endpoints were unreachable. We also report a negative result, a connector whose stored header mapping did not satisfy its provider's authentication contract, which demonstrates that centralized custody does not by itself guarantee correct configuration. We conclude that vault mediation eliminates several disclosure paths but is necessary rather than sufficient: least privilege, deterministic action authorization, human approval, telemetry redaction, and rotation remain independently required. The study is purposive and small; it is a functional security evaluation, not a certification.
  \end{abstract}
  \vspace{10pt}
  \end{@twocolumnfalse}
]

\section{Introduction}

The deployment pattern that makes LLM agents commercially useful, namely connecting them to enterprise APIs, SaaS platforms, internal services, and Model Context Protocol (MCP) servers, is the same pattern that makes them a credential-handling risk. A conventional application confines a secret to a secrets manager and a small number of trusted processes. An agentic application introduces a probabilistic planner, a persistent conversation, retrieval over untrusted documents, tool schemas, execution traces, inter-agent messages, and user-visible outputs. Each is a potential copy site for anything that enters the model's context.

The practical consequence is that the common user behavior of pasting an API key or access token directly into a chat interface is not merely poor hygiene. It is a boundary violation whose blast radius is difficult to bound after the fact, because the credential is duplicated into systems that were designed for retention and retrieval rather than for secrecy.

Existing research establishes the two halves of this problem separately. One body of work shows that instructions embedded in retrieved content can redirect an LLM application's behavior~\cite{greshake2023,perez2022,liu2024formalizing,liu2023appinjection}, and that tool-using agents inherit this exposure through every document and API response they read~\cite{toolemu,agentdojo2024,injecagent2024}. A second body of work shows that credentials are abundant in the artifacts these systems consume and produce~\cite{meli2019,pearce2022}, and that models can surface memorized secrets from their training data~\cite{carlini2019,carlini2021,nasr2023}. What has received less attention is the execution boundary between them: where the credential physically resides at the moment an agent makes an authenticated call, and what an attacker who controls the agent's instructions can therefore reach.

This paper makes four contributions:

\begin{enumerate}
  \item A structured threat chain (\S\ref{sec:threat}) describing how a credential propagates from introduction to downstream compromise in agentic systems.
  \item A synthesis of the available evidence (\S\ref{sec:evidence}), carefully separating incident disclosure, vendor-reported measurement, and security guidance.
  \item A description of a vault-mediated execution boundary (\S\ref{sec:arch}) as implemented in Corvic Security Vault, based on internally observable platform contracts.
  \item Two controlled experiments and their results (\S\ref{sec:method}--\S\ref{sec:results}), a threat-model assessment (\S\ref{sec:risk}), a deployment blueprint (\S\ref{sec:blueprint}), and an explicit account of limitations (\S\ref{sec:limits}).
\end{enumerate}

\paragraph{Scope and safety boundary.} All experiments were non-destructive and read-only with respect to third-party services. No secret value was printed, copied, transmitted to an unauthorized destination, or validated outside its configured service. The experiments tested for the \emph{presence} of credential material and for authentication \emph{behavior}; they did not attempt to recover, reconstruct, or exfiltrate any secret.

\section{Threat Model}\label{sec:threat}

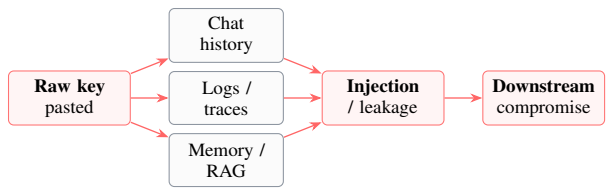
\begin{figure}[t]
\centering
\begin{tikzpicture}[
  font=\scriptsize,
  node distance=3mm,
  box/.style={draw=cvgray!70,fill=gray!3,rounded corners=2pt,align=center,inner sep=3pt,minimum height=7mm,text width=13mm},
  red/.style={draw=red!55,fill=red!4,rounded corners=2pt,align=center,inner sep=3pt,minimum height=7mm,text width=14mm},
  ar/.style={-{Stealth[length=1.6mm]},draw=red!60,line width=0.5pt}
]
\node[red] (src) {\textbf{Raw key}\\pasted};
\node[box,above right=1mm and 5mm of src] (hist) {Chat\\history};
\node[box,right=5mm of src] (log) {Logs /\\traces};
\node[box,below right=1mm and 5mm of src] (mem) {Memory /\\RAG};
\node[red,right=5mm of log] (inj) {\textbf{Injection}\\/ leakage};
\node[red,right=5mm of inj] (dwn) {\textbf{Downstream}\\compromise};
\draw[ar] (src) -- (hist);
\draw[ar] (src) -- (log);
\draw[ar] (src) -- (mem);
\draw[ar] (hist) -- (inj);
\draw[ar] (log) -- (inj);
\draw[ar] (mem) -- (inj);
\draw[ar] (inj) -- (dwn);
\end{tikzpicture}
\caption{The direct-paste anti-pattern. One introduction event creates several durable copies in systems designed for retention, any of which can later feed disclosure or misuse.}
\label{fig:chain}
\end{figure}

We model credential exposure in agentic systems as a five-stage chain. Figure~\ref{fig:chain} illustrates the propagation stage, where a single introduction event produces multiple durable copies before any API request is issued.

\begin{description}[leftmargin=0pt,style=nextline,itemsep=2pt]
  \item[S1: Introduction.] A user message, prompt template, source file, notebook, tool definition, or MCP server configuration contains a reusable secret.
  \item[S2: Propagation.] Model context, conversation history, execution logs, memory and vector stores, generated code, and error payloads duplicate it.
  \item[S3: Trigger.] Direct or indirect prompt injection, system-prompt extraction, a malicious tool response, or ordinary model error requests disclosure or misuse.
  \item[S4: Execution.] A broadly privileged token authorizes reads, writes, deletion, publication, compute consumption, or lateral movement.
  \item[S5: Persistence.] A long-lived bearer token is replayable outside the original agent, user, session, or enterprise control plane.
\end{description}

Stage S3 is the best-studied link. Indirect prompt injection was demonstrated against real LLM-integrated applications by Greshake et al.~\cite{greshake2023}, following earlier work on direct instruction override~\cite{perez2022}; Liu et al.~\cite{liu2024formalizing} subsequently formalized the attack class and benchmarked ten defenses, finding none robust. For agents specifically, InjecAgent~\cite{injecagent2024} and AgentDojo~\cite{agentdojo2024} show that tool-integrated agents remain vulnerable in realistic multi-step environments, and Fang et al.~\cite{fang2024} show that agents with tool access can carry out end-to-end intrusions autonomously. Proposed defenses operate either on the prompt boundary, by structurally separating instructions from data~\cite{chen2024struq,hines2024}, or on the execution boundary, by extracting control and data flow from the trusted query and enforcing capability policies at tool-call time~\cite{camel2025}. The architecture evaluated here is complementary to the latter: it constrains what a tool call can authenticate to, rather than which tool call is emitted.

Stage S5 has an empirical basis as well. Meli et al.~\cite{meli2019} showed that secrets in public repositories are both common and long-lived, and Pearce et al.~\cite{pearce2022} showed that code-generation models emit insecure patterns at measurable rates, which matters when an agent writes the integration code. Models can also surface secrets they memorized during training~\cite{carlini2019,carlini2021,nasr2023}, and corpora themselves can be poisoned at practical cost~\cite{carlini2023poison}.

Two properties of this chain motivate the architecture in \S\ref{sec:arch}. First, stages S1--S2 are \emph{eliminable by construction}: if no reusable credential ever enters model-visible context or agent-visible runtime state, most copy sites never receive one. Second, stages S4--S5 are \emph{not} addressed by custody alone: an agent holding no secret can still invoke an authorized but destructive operation. Any credible defense must therefore combine credential separation with independent authorization.

\section{Evidence and Related Work}\label{sec:evidence}

We deliberately separate three classes of evidence, which are frequently conflated in vendor literature: a platform incident disclosure, a vendor-reported measurement, and peer-reviewed research and standards guidance.

\subsection{Platform incident disclosure}

On 1 June 2024, Hugging Face disclosed that it had detected unauthorized access to its Spaces secrets and advised affected users to rotate credentials~\cite{hf2024}. This supports a bounded claim: a hosted AI platform detected access involving its secret store and treated configured credentials as potentially exposed. It does \emph{not} establish that every secret was exfiltrated or abused, and it is not an instance of a model spontaneously revealing a credential. Its relevance is that hosted agent platforms concentrate credentials, making rotation, revocation, scoping, and audit first-order platform requirements.

\subsection{Vendor-reported measurement}

GitGuardian reports that its detectors identified 23{,}770{,}171 hardcoded secrets in public GitHub commits made during 2024, a 25\% year-over-year increase, and that 4.61\% of the 69.6~million public repositories it scanned contained at least one secret~\cite{gg2025}. This is a vendor measurement, not a census: detector coverage, secret definitions, deduplication policy, and validity checking all affect the figure, and detection does not establish that a credential was live or exploited. Interpreted conservatively, it establishes that agent developers operate in an environment with a large pre-existing supply of exposed credentials. The academic record is consistent with this: Meli et al.~\cite{meli2019} found secrets in public GitHub repositories at scale and observed that many remained valid long after commit, which is the property that makes a leaked credential useful to an attacker.

\subsection{Security guidance}

OWASP's 2025 LLM guidance treats sensitive-information disclosure~\cite{owasp_llm02}, system-prompt leakage~\cite{owasp_llm07}, and excessive agency~\cite{owasp_llm06} as distinct but interacting risks, and recommends handling privileged functionality in code rather than exposing tokens to the model, executing tools in the user's context, and applying minimum scopes. OWASP's MCP Top 10 entry MCP01 specifically names token mismanagement and secret exposure, highlighting hard-coded credentials, long-lived tokens, secrets retained in model memory or protocol logs, and retrieval of such material through prompt recall or log inspection~\cite{owasp_mcp01}. NIST's Generative AI Profile distinguishes direct from indirect prompt injection and notes downstream consequences for interconnected systems~\cite{nist2024}; NIST's adversarial machine learning taxonomy classifies these attack classes and their mitigations~\cite{nist100_2}, and MITRE ATLAS catalogues observed techniques against AI-enabled systems~\cite{atlas}. Research on tool-using agents further motivates bounding and evaluating agent tool authority rather than trusting model output~\cite{toolemu,greenblatt2023}.

Measurement work on MCP deployments substantiates the protocol-level concern. Studies of live MCP ecosystems report over-broad API privileges granted to servers~\cite{mcpprivilege2025}, practical tool-poisoning against real servers~\cite{mcptox2025}, a taxonomy of malicious-server behaviors~\cite{mcpattack2025}, and ecosystem-wide security weaknesses~\cite{mcpmeasure2025}.

The guidance converges on a single architectural prescription: \emph{separate the planner from the credential, and authorize the action independently of the planner's prose.} This is not a new principle so much as an old one applied to a new component. Saltzer and Schroeder's least-privilege and complete-mediation principles~\cite{saltzer1975} and Hardy's confused-deputy problem~\cite{hardy1988} describe exactly the failure mode of an agent that holds broad authority and accepts instructions from an untrusted source; NIST's zero trust architecture generalizes the response~\cite{nist800207}.

\section{Vault-Mediated Execution}\label{sec:arch}

\begin{figure}[t]
\centering
\begin{tikzpicture}[
  font=\scriptsize,
  box/.style={draw=cvgray!70,fill=gray!3,rounded corners=2pt,align=center,inner sep=3pt,text width=16mm,minimum height=11mm},
  vault/.style={draw=cvgreen,fill=cvgreen!6,rounded corners=3pt,align=center,inner sep=3pt,text width=20mm,minimum height=17mm,line width=0.8pt},
  ar/.style={-{Stealth[length=1.8mm]},draw=cvgray,line width=0.6pt}
]
\node[box] (user) {\textbf{Agent}\\intent, plan\\\emph{no secret}};
\node[box,right=4mm of user] (req) {\textbf{Tool request}\\host, method,\\arguments};
\node[vault,right=4mm of req] (v) {\textbf{Vault boundary}\\connector by ID\\inject at request\\origin binding\\lifecycle state};
\node[box,right=4mm of v] (svc) {\textbf{Service}\\authorized\\call};
\draw[ar] (user) -- (req);
\draw[ar] (req) -- (v);
\draw[ar] (v) -- (svc);
\draw[ar] (svc.south) -- ++(0,-5mm) -| node[pos=0.25,below,font=\tiny]{response returns; reusable secret does not} (req.south);
\end{tikzpicture}
\caption{Vault-mediated execution. The agent names a connector; the trusted request boundary supplies authentication for the connector's configured origin. The credential is never a value the agent holds.}
\label{fig:arch}
\end{figure}
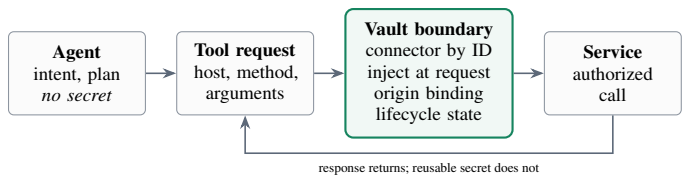

Corvic Security Vault exposes each external integration as a \emph{connector} record carrying a configured API origin, a connector type (\texttt{api}, \texttt{oauth}, or \texttt{mcp}), non-secret configuration, named secret entries with their header mappings, and lifecycle state (for example \texttt{active}, \texttt{expired}, or unconfigured). Agent-executed code receives only a list of connector identifiers, not secret values. The platform's own execution contract states that no ambient cloud credentials are available to that code and that connector credentials are injected as HTTP headers for the connector's declared host.

This yields four security properties that are testable from inside the sandbox without privileged access:

\begin{enumerate}[label=P\arabic*.,leftmargin=2.2em]
  \item \textbf{Non-observability.} Executing code cannot read the credential from its environment, its own prepared request headers, or conventional on-disk credential locations.
  \item \textbf{Origin binding.} The credential is attached to requests to the connector's configured origin and not to arbitrary destinations.
  \item \textbf{No ambient identity.} The sandbox holds no cloud workload identity that could substitute for the vault.
  \item \textbf{Centralized lifecycle.} Connector state, type, and scope metadata are managed centrally and surfaced as metadata rather than as secret values.
\end{enumerate}

\section{Experimental Method}\label{sec:method}

We evaluated P1--P4 with two black-box experiments executed inside the platform's sandboxed Python environment on 27 September 2026 (UTC). Each experiment was granted exactly one connector, and each probe carried an explicit binary expectation defined before execution.

\begin{table}[t]
\centering
\small
\caption{Experimental design. Each experiment was granted a single connector identifier and issued a fixed probe set.}
\label{tab:design}
\footnotesize
\setlength{\tabcolsep}{4pt}
\begin{tabularx}{\columnwidth}{@{}l>{\RaggedRight\arraybackslash}X@{}}
\toprule
\multicolumn{2}{@{}l}{\textbf{E1: runtime exposure} \quad connector: FRED} \\
\multicolumn{2}{@{}l}{\footnotesize declared origin \textsf{api.stlouisfed.org}} \\
\midrule
Probe set & Environment variable names and values; the process environment file; 12 well-known credential paths; secret-mount globs; AWS IMDSv1 and GCP metadata endpoints; declared-host call; three echo services; two unrelated APIs \\
\midrule
\multicolumn{2}{@{}l}{\textbf{E2: host-bound function} \quad connector: GitHub} \\
\multicolumn{2}{@{}l}{\footnotesize declared origin \textsf{api.github.com}} \\
\midrule
Probe set & Authenticated user endpoint; caller-prepared request headers; credential pattern scan over the environment; two echo services; one unrelated authenticated API \\
\bottomrule
\end{tabularx}
\end{table}

\paragraph{Controls.} The design pairs a \emph{positive control} (an authenticated call to the connector's declared origin must succeed) with \emph{negative controls} (echo services must not receive an \texttt{Authorization} header; an unrelated authenticated API must report missing authentication). Without both, a passing result would be indistinguishable from a sandbox with no network access.

\paragraph{Detection.} Credential presence was tested with high-confidence prefix patterns for common token formats and with keyword matching over environment variable names. Echo-service responses were searched for authorization headers and token-shaped strings. Detection results are reported as counts and booleans only.

\paragraph{Classification.} Each probe was scored against its pre-declared expectation and assigned to one of seven control domains. Connector \emph{functionality} was scored separately from security controls, so that a provider-contract failure could not be recorded as a security pass or a security failure.

\section{Results}\label{sec:results}

\begin{figure*}[t]
\centering
\includegraphics[width=0.94\textwidth]{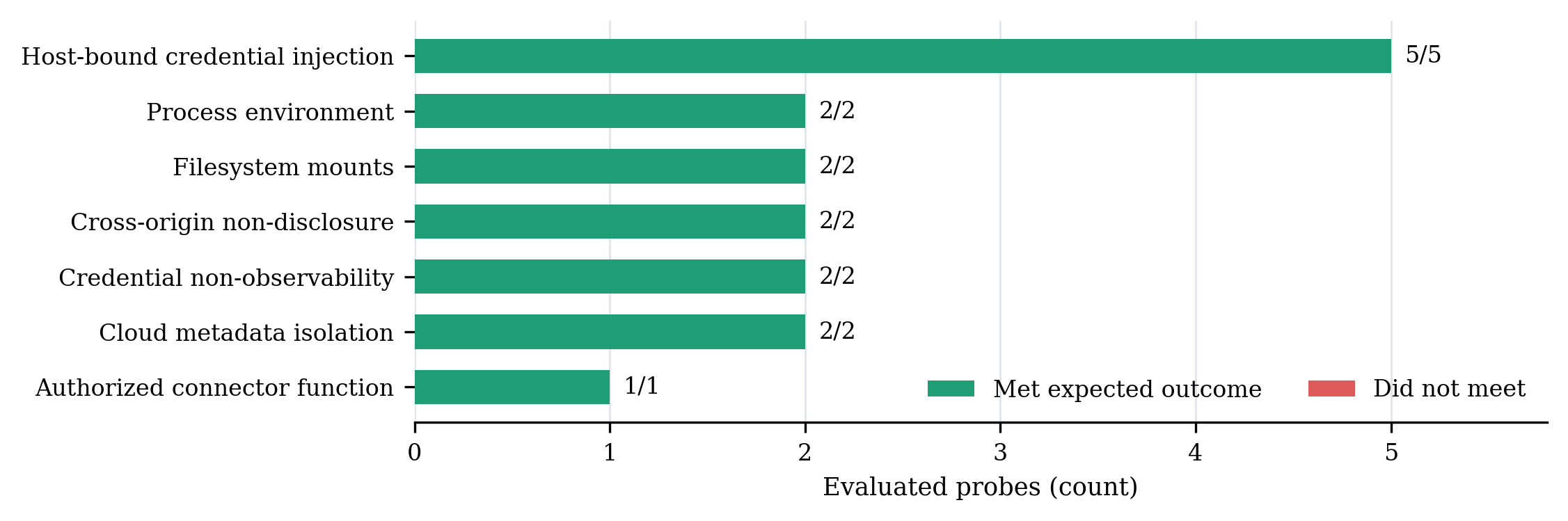}
\caption{Evaluated probes by control domain, from experiments E1 and E2. Every probe met its pre-declared expected outcome. The result is a controlled functional check on a small purposive sample, not a statistical pass-rate estimate.}
\label{fig:probes}
\end{figure*}

All 16 evaluated probes met their expected security outcome (Table~\ref{tab:results}, Figure~\ref{fig:probes}).

\begin{table}[t]
\centering
\small
\caption{Evaluated probes by control domain. Rates are reported for completeness; per-domain $n$ ranges from 1 to 5 and no confidence interval is meaningful for this purposive design.}
\label{tab:results}
\begin{tabular}{@{}lrrr@{}}
\toprule
\textbf{Control domain} & $n$ & \textbf{Pass} & \textbf{Fail} \\
\midrule
Host-bound credential injection & 5 & 5 & 0 \\
Process environment & 2 & 2 & 0 \\
Filesystem mounts & 2 & 2 & 0 \\
Cross-origin non-disclosure & 2 & 2 & 0 \\
Credential non-observability & 2 & 2 & 0 \\
Cloud metadata isolation & 2 & 2 & 0 \\
Authorized connector function & 1 & 1 & 0 \\
\midrule
\textbf{Total} & \textbf{16} & \textbf{16} & \textbf{0} \\
\bottomrule
\end{tabular}
\end{table}

\subsection{Positive control (P2, functional)}

In E2, an authenticated \textsf{GET} of the GitHub user endpoint (\textsf{api.github.com/user}) returned HTTP~200 with an authenticated identity while the executing module supplied no \texttt{Authorization} header and contained no credential. This establishes that the vault performed authentication for the connector's declared origin.

\subsection{Non-observability (P1)}

The sandbox exposed 20 environment variables. None had a secret-suggestive name, and none matched the high-confidence credential patterns tested. The process environment file \textsf{/proc/self/environ} was readable (634 bytes) but contained no matching secret. None of the 12 tested credential paths existed, and secret-mount globs returned no entries. A \texttt{requests} object prepared by the calling process for the GitHub endpoint carried no \texttt{Authorization} header, indicating that injection occurs outside the process's own view of the request.

\subsection{Origin binding and cross-origin non-disclosure (P2)}

Three third-party echo services (httpbin, Postman Echo, httpbingo) returned HTTP~200 and reflected no connector \texttt{Authorization} header and no token-shaped string; the reflected requests identified the caller only as the platform sandbox. Calls to the OpenAI model-listing endpoint (\textsf{api.openai.com/v1/models}) returned HTTP~401 with a missing-authentication error in both experiments, including while a GitHub connector was attached. The credential was therefore neither broadcast to arbitrary destinations nor reused across origins.

\subsection{Ambient identity (P3)}

Requests to the GCP metadata service-account token endpoint and to AWS IMDSv1 both timed out after 6~s. Within this run, no cloud workload credential was reachable. We note that a timeout demonstrates non-reachability during the run rather than constituting a formal network-isolation proof.

\subsection{Negative result: configuration is not custody}

In E1, the declared-host call to \textsf{api.stlouisfed.org} returned HTTP~400 with the message that the \texttt{api\_key} variable was not set. The connector's stored secret entry was mapped to an \texttt{Authorization} header, whereas the provider requires a query parameter. This is a functional failure with a secure outcome: no credential was disclosed, and no unauthorized access occurred. It is reported because it isolates an important limit. Centralizing custody does not validate each provider's authentication contract, and a misconfigured connector fails closed but also fails to work.

\section{Threat-Model Assessment}\label{sec:risk}

\begin{figure*}[t]
\centering
\includegraphics[width=0.82\textwidth]{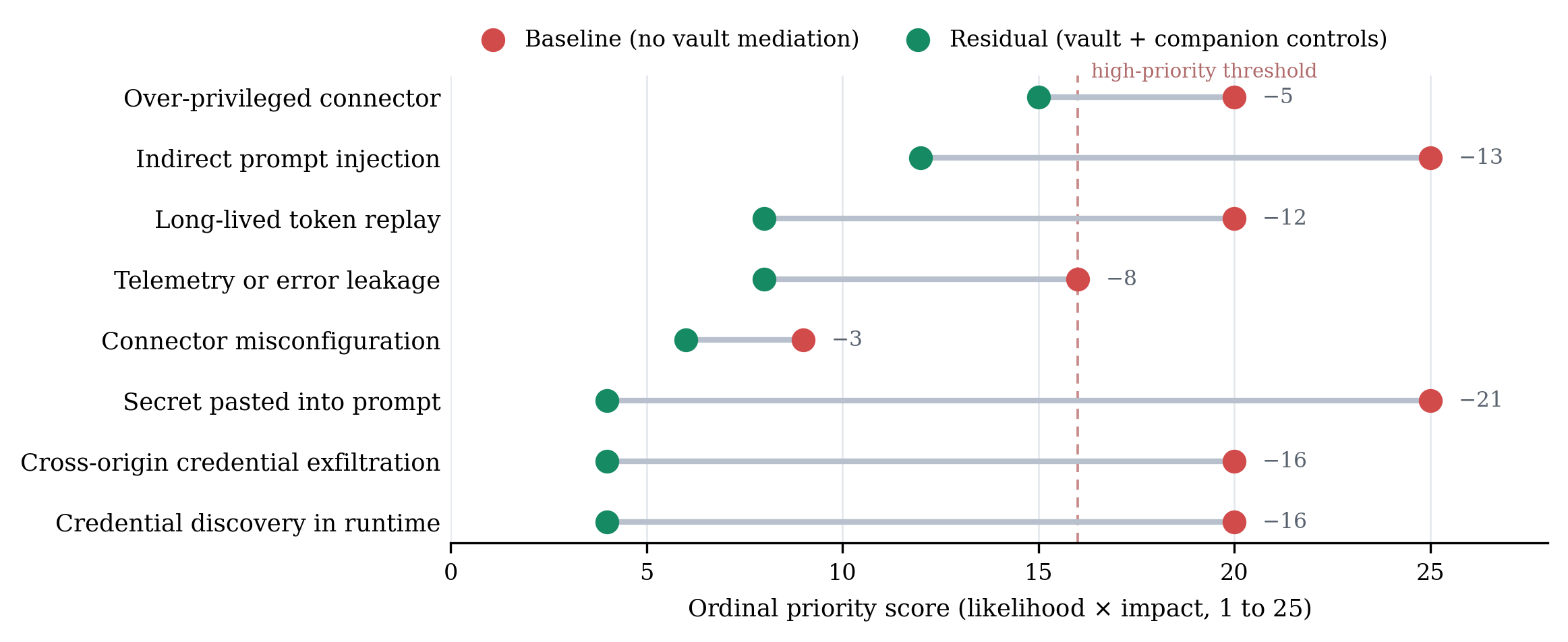}
\caption{Assessed change in ordinal priority score for the eight threats of Table~\ref{tab:risk}, sorted by residual score. Each pair joins the baseline score to the residual score after vault mediation, with the reduction annotated at right. Three threats remain at or near the high-priority threshold, so credential custody alone does not retire them.}
\label{fig:risk}
\end{figure*}

Table~\ref{tab:risk} and Figure~\ref{fig:risk} summarize a directional assessment of eight threats before and after vault mediation. Ratings are ordinal expert judgments on a 1--5 scale, not calibrated probabilities; the product $L\times I$ is used only for prioritization.

\begin{table}[t]
\centering
\footnotesize
\caption{Directional threat assessment. $L{\times}I$ is an ordinal prioritization score, not a probability.}
\label{tab:risk}
\begin{tabularx}{\columnwidth}{@{}r>{\RaggedRight}Xrr@{}}
\toprule
\textbf{\#} & \textbf{Threat} & \textbf{Base} & \textbf{Resid.} \\
\midrule
1 & Secret pasted into prompt & 25 & 4 \\
2 & Indirect prompt injection & 25 & 12 \\
3 & Cross-origin credential exfiltration & 20 & 4 \\
4 & Credential discovery in runtime & 20 & 4 \\
5 & Over-privileged connector & 20 & 15 \\
6 & Long-lived token replay & 20 & 8 \\
7 & Telemetry or error leakage & 16 & 8 \\
8 & Connector misconfiguration & 9 & 6 \\
\bottomrule
\end{tabularx}
\end{table}

Two threats move least. \emph{Over-privileged connector} is barely reduced, because vault custody does not narrow provider-side scopes: a valid token with write authority remains a valid token with write authority. \emph{Indirect prompt injection} is reduced but not resolved, because the attack targets the agent's action selection rather than the secret itself, and no published defense fully prevents it~\cite{liu2024formalizing,agentdojo2024}. Both residuals must be addressed by least privilege and by authorization outside the model.

\section{Deployment Blueprint}\label{sec:blueprint}

\begin{enumerate}[leftmargin=1.4em,itemsep=2pt]
  \item \textbf{Inventory.} Record owner, environment, provider, authentication type, data classification, scopes, expiry, and rotation path per connector.
  \item \textbf{Eliminate direct paste.} Detect and block credential patterns in chat input, prompts, notebooks, tool schemas, and logs; offer a one-click connector path instead.
  \item \textbf{Minimize provider authority.} Separate read and write identities, prefer per-user OAuth over shared administrator tokens, and bind audience and resource where supported. OAuth's own security guarantees depend on deployments following its recommendations~\cite{fett2016}; treat delegated tokens as narrow capabilities~\cite{hardy1988}, not as ambient authority.
  \item \textbf{Authorize outside the model.} Validate host, method, path, schema, tenant, volume, and user entitlement at a deterministic policy point, as complete mediation requires~\cite{saltzer1975,nist800207}. Control-flow and capability approaches such as CaMeL~\cite{camel2025} show this can be enforced systematically rather than heuristically.
  \item \textbf{Require approval.} Gate transfers, deletion, external publication, privilege change, bulk export, production mutation, and high-cost calls.
  \item \textbf{Control egress.} Allow-list destinations and prevent redirects or tool output from relocating credentials or sensitive results.
  \item \textbf{Redact telemetry.} Strip authorization headers, cookies, query secrets, and signed URLs; retain connector identifier, policy decision, and correlation identifiers.
  \item \textbf{Exercise lifecycle.} Test OAuth refresh and expiry, revocation, connector disablement, offboarding, incident rotation, and downstream audit review.
  \item \textbf{Red-team continuously.} Cover direct and indirect injection, tool poisoning, cross-origin requests, error paths, retries, redirects, memory recall, and multi-agent handoff. Published agent benchmarks~\cite{agentdojo2024,injecagent2024,mcptox2025} supply reusable test cases.
\end{enumerate}

\section{Limitations}\label{sec:limits}

The evaluation is black-box and purposive. Sixteen probes cannot support prevalence estimates, and the design includes no random sampling, no repeated-trial variance, and no concurrency stress. We did not test redirect chains, DNS rebinding, OAuth refresh capture, MCP session state, browser-tool paths, or adversarial compromise of a configured destination. Pattern-based detection can miss unknown credential formats. Network timeouts evidence non-reachability during the run rather than a formal isolation guarantee. The positive control exercises one connector, not every connector.

The following properties were \textbf{not verified} and are not claimed: encryption algorithms and key management; token-at-rest storage; tenant-isolation internals; audit-log completeness; administrative access controls; deletion guarantees; rotation service levels; compliance certifications; and request-policy semantics beyond the tested hosts.

Future work should add provider-contract conformance tests, explicit policy-denial tests, redirect and egress handling, per-tenant isolation verification, audit-log review, rotation timing measurement, and independent penetration testing. Running an established agent-injection benchmark~\cite{agentdojo2024,injecagent2024} against a vault-mediated deployment would measure the residual in Table~\ref{tab:risk} rather than assess it, and is the most valuable single extension of this work.

\section{Conclusion}

The enterprise question is not whether an LLM can be instructed to keep a secret; it cannot be trusted as an enforcement point. The correct design keeps reusable credentials outside the model's vocabulary and outside agent-visible runtime state, then mediates every authenticated action through deterministic identity, policy, and audit controls. In the configuration tested, Corvic Security Vault implemented a strong version of that separation: agent code named a connector and completed an authenticated service call without ever holding a reusable credential, while unrelated origins received none. The residual agenda of least privilege, action authorization, human approval, redaction, rotation, configuration validation, and continuous verification is where the remaining risk lives.

\section*{Reproducibility and Ethics}

Both experiments were executed in the authors' own tenant against connectors configured for that tenant. All third-party requests were read-only. No secret value was printed, stored, transmitted to an unauthorized destination, or tested against any service other than its configured provider. Results are reported as counts, status codes, and booleans. No third-party system was attacked, and no vulnerability in any external provider was sought or exercised.

\end{document}